\documentstyle[preprint,eqsecnum,aps]{revtex}
\def\bra#1{\langle #1 |}
\def\ket#1{| #1 \rangle}
\def\inner#1#2{\langle #1|#2 \rangle}
\begin{document}
\draft
\title{The Decoherence of Phase Space Histories}
\author{Todd A. Brun}
\address{Department of Physics, Caltech, Pasadena, CA  91125}
\date{\today}

\maketitle

\begin{abstract}
In choosing a family of histories for a system, it is often convenient to
choose a succession of locations in phase space, rather than configuration
space, for comparison to classical histories.  Although there are no good
projections onto phase space, several approximate projections have been used
in the past; three of these are examined in this paper.  Expressions are
derived for the probabilities of histories containing arbitrary numbers of
projections into phase space, and the conditions for the decoherence of these
histories are studied.
\end{abstract}


\section{Introduction}

A great deal of work has been done recently on
the use of the decoherence formalism
to describe quantum mechanical systems
\cite{GMHart1,GMHart2,GMHart3,Omnes,Zurek1,Zurek2,DowkHall,Brun1}.
These systems can be described in terms of
{\it decoherent histories}, which
can be assigned probabilities obeying classical probability sum rules.
While, in principle, a history could be described
in terms of any set of variables, most of the
work has focussed on histories of particles in configuration space.
The simplest and most fine-grained such history is just the classical
trajectory of a particle, specifying its exact position at every moment
in time.  Such histories do not decohere, however.  Instead, one must
consider considerably coarse-grained histories, in which a position is
given only at certain discrete times, and only within certain finite
intervals.  A history can then be specified by a string of indices
$\alpha_i$, stating which interval the particle is in at time $t_i$.

Another important class of histories, though, would be descriptions of a system
as being in cells of phase space at successive points in time.  A small amount
of work has been done on this subject
\cite{Brun2,Halliwell1,Halliwell2,Halliwell3,Twamley1},
but they have not been tackled in full
generality.

\subsection{The Decoherence Functional}

The decoherence functional is a functional on {\it pairs of histories} of
a quantum mechanical system.  One simple description of the functional (though
not the most general) has the form:
\begin{equation}
D[\alpha,\alpha^\prime] = {\rm Tr} \biggl\{ P^n_{\alpha_n}(t_n) \cdots
  P^1_{\alpha_1}(t_1) \rho P^1_{\alpha^\prime_1}(t_1)
  \cdots P^n_{\alpha^\prime_n}(t_n) \biggr\}.
\end{equation}
In this expression, $\rho$ is the initial density matrix of the system.
The $P^i_{\alpha_i}(t_i)$ are Heisenberg {\it projection operators} onto
Hilbert
space.  At each time $t_i$ these projection operators represent different
alternatives for the system.  In terms of Schr\"odinger projections $P$,
these time dependent projections can be written $P(t) = e^{iHt/\hbar} P
e^{-iHt/\hbar}$.
A complete set of histories has an exhaustive
set of alternatives at each time,
\begin{equation}
\sum_{\alpha_i} P^i_{\alpha_i}(t_i) = 1.
\end{equation}
A particular choice of the $\{\alpha_i\}$
represents one particular history, which
we will denote $\alpha$ for brevity.  Thus, saying that a given history
$\alpha$
occurs implies that alternative $\alpha_1$ occurs at time $t_1$, $\alpha_2$ at
time $t_2$, and so forth.

The probability of a given history occuring is just given by the diagonal
elements of $D[\alpha,\alpha^\prime]$:
\begin{equation}
p(\alpha) = D[\alpha,\alpha].
\end{equation}
In order for these histories to obey the classical probability sum rules, we
must require that the set of histories {\it decoheres}.  The usual requirement
for this is that the off-diagonal terms of the decoherence function vanish,
\begin{equation}
D[\alpha,\alpha^\prime] = 0,\ \ \  \alpha \ne \alpha^\prime.
\end{equation}
This is actually a sufficient, but not a necessary condition for decoherence.
All that is truly required is that the {\it real parts} of the off-diagonal
terms vanish.  Most physically decoherent systems, however, display this
stronger form of decoherence; we will see this in the cases that we consider.

\subsection{The Transition Matrix}

The most common type of problem treated at present is that in which the
variables are divided into a {\it system} and a {\it reservoir}, or
{\it environment}.  In this case one traces over the reservoir variables,
and is left with a reduced density matrix on only the system variables.
Instead of the simple time evolution operator $e^{-iHt/\hbar}$, the system
evolves according to a somewhat more complicated {\it transition matrix}
or {\it propagator} ${\bf T}$.  In terms of path integrals this is
\begin{equation}
T(x_f,x^\prime_f,t_f; x_i,x^\prime_i,t_i) = \int \delta x \delta x^\prime\
  \exp {i\over\hbar} \biggl\{ S[x(t)] - S[x^\prime(t)] +
  W[x(t),x^\prime(t)] \biggr\},
\label{propagator}
\end{equation}
where the integral is over all paths $x(t)$ and $x^\prime(t)$ from
$t_i$ to $t_f$ which begin at $x_i$ and $x^\prime_i$ and end at
$x_f$ and $x^\prime_f$, respectively.  $S[x(t)]$ is the action of the
system variables independent of the reservoir, and $W[x(t),x^\prime(t)]$
is the Feynman-Vernon {\it influence phase} arising due to the interactions
with the reservoir \cite{FeynVern}.
The model most commonly considered is one that was
developed in the study of Brownian motion \cite{CaldLegg},
in which a one-dimensional
particle described by a single variable $x$ interacts with an infinite bath
of harmonic oscillators via a linear or weakly nonlinear potential, but
this formalism is quite general.

The reduced density matrix $\rho(x; x^\prime)$ evolves straightforwardly:
\begin{equation}
\rho(x; x^\prime) |_{t_f} = \int dx_i dx^\prime_i\
  T(x_f,x^\prime_f,t_f; x_i,x^\prime_i,t_i) \rho(x_i; x^\prime_i) |_{t_i}.
\end{equation}
Thus, the decoherence functional can now be written
\begin{equation}
D[\alpha,\alpha^\prime] = Tr_x \biggl\{ P^n_{\alpha_n} {\bf T} ( \cdots
  {\bf T} ( P^1_{\alpha_1} \rho P^1_{\alpha^\prime_1} )
  \cdots ) P^n_{\alpha^\prime_n} \biggr\}.
\end{equation}

With the projections $P$ being onto intervals of coordinate space, it is
very easy to write the decoherence functional as a constrained path integral
over $x$ and $x^\prime$.  For phase-space projections, the form of the
decoherence function is more complicated, as we shall see.  While there are
no true projectors onto cells of phase space (as there are for intervals
of coordinate space), there are a number of approximate projectors, and we
shall consider these one at a time.

In dealing with phase space, it is natural to consider other representations
of the density matrix, most obviously the Wigner distribution:
\begin{equation}
w(X,p) = {1\over\pi} \int d\xi\ {\rm e}^{i\xi p/\hbar}
  \rho(X + \xi/2; X - \xi/2).
\label{wigner}
\end{equation}
This distribution acts in many ways like a probability distribution in
phase space, with the major exception that it can be negative in localized
regions.  The time evolution of $w(X,p)$ is also described by a transition
matrix:
\begin{equation}
w(X_f,p_f)|_{t_f} = \int\int dX_i dp_i\ T_w (X_f,p_f,t_f; X_i,p_i,t_i)
  w(X_i,p_i)|_{t_i}.
\end{equation}
\begin{eqnarray}
T_w(X_f,p_f,t_f; X_i,p_i,t_i) && = \int\int d\xi_f d\xi_i\
  {\rm e}^{i(p_f\xi_f - p_i\xi_i)/\hbar} \nonumber\\
&& \times T(X_f + {\xi_f\over2}, X_f - {\xi_f\over2},
  t_f; X_i + {\xi_i\over2}, X_i - {\xi_i\over2}, t_i),
\label{Tw}
\end{eqnarray}
where this $T$ is the same transition matrix defined above (\ref{propagator}).
We shall see that
the expressions for the probabilities of phase space histories are described
very naturally in terms of Wigner distributions.

\section{Probabilities of Phase Space Histories}

As mentioned above, there are no true projections onto cells of phase space
\cite{Halliwell1}.
This is essentially a side-effect of the uncertainty principle, which
prevents both $x$ and $p$ from being localized simultaneously.  However,
for cells larger in area that $\hbar$, we can approximate projections
reasonably
well.

In making these calculations, we will find that it is useful to work in
terms of $w(X,P)$ and ${\bf T_w}$.  For this we use the inverses of
(\ref{wigner}) and (\ref{Tw}), namely
\begin{equation}
\rho(x; x^\prime) = \int dp\ {\rm e}^{-ip(x-x^\prime)/\hbar}
  w( (x+x^\prime)/2, p)
\label{wigner_inverse}
\end{equation}
and
\begin{eqnarray}
T(x_f,x^\prime_f,t_f; x_i,x^\prime_i,t_i) && = {1\over\pi} \int\int dp_i dp_f\
  {\rm e}^{ i p_i (x_i - x^\prime_i) - i p_f (x_f - x^\prime_f)} \nonumber\\
&& \times T_w ( (x_f + x^\prime_f)/2, p_f, t_f;
  (x_i + x^\prime_i)/2, p_i, t_i).
\label{Tw_inverse}
\end{eqnarray}

\subsection{Consecutive $X$ and $P$ Projections}

While there are no good projections onto phase space cells, projections onto
intervals in coordinate or momentum space are perfectly well-defined and
straightforward.  They are just
\begin{mathletters}
\begin{eqnarray}
P_{x_i} && = \int_{x_i - \Delta x/2}^{x_i + \Delta x/2} dx\ \ket{x}\bra{x}, \\
P_{p_i} && = \int_{p_i - \Delta p/2}^{p_i + \Delta p/2} dp\ \ket{p}\bra{p}.
\end{eqnarray}
\label{exact_projections}
\end{mathletters}
We can imagine using a projection to determine which interval of $x$ the
system is in, followed a short time later by a projection onto an interval
of $p$.  If we let the time between these two ``measurements'' go to zero,
we can make use of the relation
\begin{equation}
\inner{x}{p} = {\rm e}^{ipx/\hbar}.
\end{equation}

This is unsatisfactory in a number of ways.  The most obvious is that $x$
and $p$ are non-commuting variables, so that $P_x$ and $P_p$ are also
non-commuting.  The order in which one makes these measurements matters,
particularly if the intervals are fairly small (compared to $\hbar$).  If we
are interested in fairly large cells in phase space, this is of less
importance; for classical and quasiclassical systems this is often the case.

Measurements of this type were treated by Halliwell \cite{Halliwell1}.
He also considered
another type of two-projection measurement: a pair of successive position
measurements, separated by a small time interval $\Delta t$, with the momentum
determined by the {\it time of flight} between $x_1$ and $x_2$.  I have
not considered this type of measurement, as it is ill-defined as $\Delta t
\rightarrow 0$, and therefore requires non-trivial consideration of the
system's time evolution between the two position projections.  For a system
with complex dynamics this is difficult.

The exact projections (\ref{exact_projections}) used above are less convenient
for the purposes of calculation, though they are certainly more correct.
For ease of computation, therefore, it is customary to use
approximate Gaussian projections,
\begin{mathletters}
\begin{eqnarray}
P_{\bar x_i} && = {1\over\sqrt\pi\Delta x} \int_{-\infty}^{\infty} dx\
  {\rm e}^{-(x- \bar x_i)^2/\Delta x^2}
  \ket{x}\bra{x}, \\
P_{\bar p_i} && = {1\over\sqrt\pi\Delta p} \int_{-\infty}^{\infty} dp\
  {\rm e}^{-(p- \bar p_i)^2/\Delta p^2}
  \ket{p}\bra{p}.
\label{approximate_projections}
\end{eqnarray}
\end{mathletters}

Using these projections, the probabilities just reduce to a product of
Gaussian integrals, which can, with a little algebra, be easily solved.
Using the expressions (\ref{wigner_inverse}) and (\ref{Tw_inverse})
for a history with $N$ measurements of cells of phase space
centered on $({\bar x}_1, {\bar p}_1), ({\bar x}_2,{\bar p}_2),
\ldots ({\bar x}_N, {\bar p}_N)$, we get
\begin{eqnarray}
P_\alpha = && {1\over{(\pi \Delta x \Delta p)^{2N}}}
  \int d\{X_i\} d\{p_i\} d\{z_i\} d\{k_i\} d\{z^\prime_i\}
  d\{k^\prime_i\} d\{\xi_i\} \nonumber\\
&& \times W(X_0,p_0)
  \delta(z_1 - (X_0 + \xi_0/2)) \delta(z^\prime_1 - (X_0 - \xi_0/2))\nonumber\\
&& \times \exp\biggl[
  - {{(z_1 - {\bar x}_1)^2 + (z^\prime_1 - {\bar x}_1)^2}\over{\Delta x^2}}
  - {{(k_1 - {\bar p}_1)^2 + (k^\prime_1 - {\bar p}_1)^2}\over{\Delta p^2}}
  \nonumber\\
&& + {i\over\hbar} \biggl( (k^\prime_1 z^\prime_1 - k_1 z_1)
  + X_1 (k_1 - k^\prime_1) + \xi_1 (p_1 - k_1/2 - k^\prime_1/2)
  - \xi_0 p_0 \biggr) \biggr] \nonumber\\
&& \times T(X_2, p_2, t_2; X_1,p_1, t_1) \\
\label{Gaussian}
&& \times \cdots \nonumber
\end{eqnarray}
The integrals over $z$, $z^\prime$, $k$, $k^\prime$, and $\xi$ are all
simple, and yield
\begin{eqnarray}
P_\alpha && = {1\over{(2\pi)^{N-1}}} \int d\{X_i\} d\{p_i\} w(X_0,p_0)
  \nonumber\\
&& \times \exp\biggl[ - {2\over{\Delta p^2}} (p_1 - {\bar p}_1)^2
  - {2\over{\Delta x^2}} (X_0 - {\bar x}_1)^2
  - {{\Delta x^2}\over{2\hbar^2}} (p_0 - p_1)^2
  - {{\Delta p^2}\over{2\hbar^2}} (X_0 - X_1)^2 \biggr] \nonumber\\
&& \times T(X_2,p_2,t_2; X_1, p_1, t_1) \nonumber\\
&& \times \exp\biggl[ - {2\over{\Delta p^2}} (p_3 - {\bar p}_2)^2
  - {2\over{\Delta x^2}} (X_2 - {\bar x}_2)^2
  - {{\Delta x^2}\over{2\hbar^2}} (p_2 - p_3)^2
  - {{\Delta p^2}\over{2\hbar^2}} (X_2 - X_3)^2 \biggr] \nonumber\\
&& \times T(X_4,p_4,t_3; X_3, p_3, t_2) \nonumber\\
&& \times \cdots \nonumber\\
&& \times \exp\biggl[ - {2\over{\Delta x^2}} (X_{2N-2} - {\bar x}_N)^2
  -{{\Delta x^2 (p_{2N-2} - {\bar p}_N)^2}\over
    {2[\hbar^2 + \Delta x^2 \Delta p^2/4]}} \biggr].
\end{eqnarray}

Note that the expression for the probability behaves very reasonably,
i.e., the evolution after a ``measurement'' continues to be centered
about the measured values of $X$ and $p$, with a spread determined by the
size of the phase space cell.

A measurement of $p$ followed by a measurement of $X$ produces an expression
very similar to the above, and is readily evaluated by the same methods.
The differences are chiefly notable when the cell size is small compared to
$\hbar$.

\subsection{Coherent State Projections}

The closest thing to a true projection onto
a cell in phase space is probably the
{\it coherent state} projection $\ket{{\bar p},{\bar x}}\bra{{\bar p},{\bar
x}}$
centered on $({\bar p},{\bar x})$.  While these are true projections, they
are not orthogonal:
\begin{equation}
|\inner{{\bar p},{\bar x}}{{\bar p^\prime},{\bar x^\prime}}|^2
  = \exp[ - ({\bar x} - {\bar x^\prime})^2/2\sigma
  - \sigma ({\bar p} - {\bar p^\prime})^2/2\hbar^2 ].
\end{equation}
Also, these states are overcomplete.  Thus, phase space histories built from
coherent states cannot be truly decoherent, and can only be even approximately
decoherent if a discrete sample of them (e.g., the states corresponding to
a lattice of points in phase space) is taken.  In a coordinate basis we
can represent a coherent state as
\begin{equation}
\inner{x}{{\bar p},{\bar x}} = {1\over{(\pi\sigma)^{1/4}}} \exp\biggl[
  - {{(x - {\bar x})^2}\over{2\sigma}} + {{i {\bar p} x}\over{\hbar}} \biggr].
\end{equation}
This expression is useful in evaluating the probability of a coherent state
history.

When we consider a history of $N$ ``measurements'' in phase space using
coherent state projections we get an expression analagous to (\ref{Gaussian}),
which can (again) be solved for the probability:
\begin{eqnarray}
P_\alpha && = {4^N\over{2\pi\sigma}} \int d\{X_i\} d\{p_i\}\
  w(X_0,p_0) \nonumber\\
&& \times \exp\biggl[ - {(X_0 - {\bar x}_1)^2\over\sigma}
  - {(X_1 - {\bar x}_1)^2\over\sigma}
  - {{(p_0 - {\bar p}_1)^2 \sigma}\over{\hbar^2}}
  - {{(p_1 - {\bar p}_1)^2 \sigma}\over{\hbar^2}} \biggr] \nonumber\\
&& \times T(X_2, p_2, t_2; X_1, p_1, t_1) \nonumber\\
&& \times\cdots \nonumber\\
&& \times T(X_{2N-2}, p_{2N-2}, t_N; X_{2N-3}, p_{2N-3}, t_{N-1}) \nonumber\\
&& \times \exp\biggl[ - { (X_{2N-2} - {\bar x}_N)^2\over\sigma}
  - {{(p_{2N-2} - {\bar p}_N)^2 \sigma}\over{\hbar^2}} \biggr].
\end{eqnarray}

The general behavior of the probabilities is very similar to that in the
first case we considered, but even cleaner and easier to see.  Coherent states
are an excellent way of representing phase space histories.

There is one other kind of approximate projection that we could consider.
It is not, in my opinion, a very attractive one, but it has been used in
the literature, and so might as well be treated here.  Consider
approximate ``projections'' of the form
\begin{equation}
P_{({\bar p},{\bar x})} = {1\over{\pi\Delta p\Delta x}} \int dp dx\
  \ket{p,x} {\rm e}^{ - (p - {\bar p})^2/\Delta p^2
    - (x - {\bar x})^2/\Delta x^2 } \bra{p,x}.
\end{equation}
If we consider $N$ measurements of this form, the probability becomes
\begin{eqnarray}
P_\alpha = && { { 4\pi\sigma\Delta p \Delta x }\over{ 2\sigma +
  \sigma^2\Delta p^2 + \Delta x^2 + \sigma\Delta p^2 \Delta x^2 }}
  \int d\{X_i\} d\{p_i\}\ w(X_0,p_0) \nonumber\\
&& \times \exp\biggl[ - { { (X_0 - {\bar x}_1)^2 + (X_1 - {\bar x}_1)^2 }
  \over{ \sigma + \Delta x^2 }}
  - \sigma {{ (p_0 - {\bar p}_1)^2 + (p_1 - {\bar p}_1)^2 }
  \over{\hbar^2  + \sigma \Delta p^2 }} \nonumber\\
&& - (\sigma^2\Delta p^2/\hbar^2 + \Delta x^2
  + \sigma\Delta p^2\Delta x^2/\hbar^2)
  \biggl( {{ (X_0 - X_1)^2 }\over{ 2\sigma(\sigma + \Delta x^2) }}
  +  {{ (p_0 - p_1)^2 }\over{ 2(\hbar^2 + \sigma\Delta p^2) }}
  \biggr)\biggr] \nonumber\\
&& \times T(X_2, p_2, t_2; X_1, p_1, t_1) \nonumber\\
&& \times \cdots \nonumber\\
&& \times T(X_{2N-2},p_{2N-2},t_N; X_{2N-3},p_{2N-3},t_{N-1})\nonumber\\
&& \times \exp\biggl[
  - 2{{ \sigma(\sigma + \Delta x^2)(p_{2N-2} - {\bar p}_N)^2
  + (\hbar^2 + \sigma\Delta p^2)(X_{2N-2} - {\bar x}_N)^2 }\over
  { 2\sigma\hbar^2 + \sigma^2\Delta p^2 + \Delta x^2\hbar^2 +
  \sigma\Delta p^2\Delta x^2 }} \biggr].
\end{eqnarray}
Again, the same sort of qualitative behavior, but a much uglier expression.

\section{Decoherence of Phase Space Histories}

While the above expressions are highly intuitive in their qualitative
behavior, we have (in a sense) been putting the cart before the horse.  It is
meaningless to assign a probability to a history without first being assured
that the set of histories described is decoherent.  There is nothing in
the expressions above to prevent one from choosing extremely tiny cells in
phase space, with areas small compared to $\hbar$; yet such histories are
certainly not decoherent, as they flagrantly violate the uncertainty
principle.

Unfortunately, while we can write {\it expressions}
for the probabilities without
having to know much about the physics of the system (i.e., the actual
behavior of the transition matrix ${\bf T}$), in order to
actually calculate them, or to say much about
decoherence, we need to know something about the path integrals.

Except in the case of quadratic systems, these integrals are not exactly
solvable.  Limited treatments of this case have been considered elsewhere
\cite{Twamley1,Twamley2}.
Most interesting systems, however, include nonlinearities.  This can be handled
in one of three ways:  numerically; in perturbation theory; or in the
semiclassical limit, where solutions are peaked about the ``classical
trajectory.''  The first approach is robust, but does not lend
itself to general arguments.  While almost any approach will eventually
have to be treated numerically to calculate actual values for probabilities
or decoherence functional elements, one would hope to get a rough idea as
to a systems behavior before invoking that numerical machinery.
The second approach, perturbation theory, is the most commonly
adopted.  When the nonlinearities are weak, the path integrals can be
approximated with considerable precision.  Unfortunately, many interesting
cases (e.g., chaotic systems) cannot be treated in this fashion; for them,
their nonlinearities are intrinsically important.  The last approach is limited
to systems with sufficient mass and inertia to resist quantum
fluctuations \cite{GMHart3}.
This is useful in considering either the classical limit
of quantum systems, or in estimating quantum effects in otherwise classical
systems, and is the approach we will adopt here.

Earlier work has concentrated on distinguished systems interacting with
a large reservoir or environment whose degrees of freedom can be neglected.
As has been shown, such systems give rise to decoherence functionals with
probabilities peaked about classical trajectories.  The transition matrix
for such a system has the form (in the limit of a large thermal reservoir)
\begin{eqnarray}
T(X_1, \xi_1, t_1; && X_0, \xi_0, t_0) = \nonumber\\
&& \int\delta X \delta\xi\ \exp{i\over\hbar}\biggl\{
  - \int_{t_0}^{t_1} \biggl( M {\ddot X}(t) + dV/dX(X(t)) + 2M\gamma{\dot X}(t)
    - g(t) \biggr) \xi(t) dt \nonumber\\
&& + {i M\gamma kT\over\hbar} \int_{t_0}^{t_1} \xi^2(t) dt
  + M \xi_1 {\dot X}_1 - M \xi_0 {\dot X}_0 + O(\xi^3) \biggr\},
\end{eqnarray}
where $X$ and $\xi$ are variables defined by
\begin{mathletters}
\begin{eqnarray}
X && = {1\over2}(x + x^\prime), \\
\xi && = x - x^\prime,
\end{eqnarray}
\end{mathletters}
and the reservoir temperature is $T$.  This is basically a toy system,
consisting of a single one-dimensional particle of mass $M$ moving in
an arbitrary potential $V(X)$.  The interaction with the reservoir provides
the dissipative term and a thermal noise; it is this noise which causes
the system to decohere.  $|\xi|$ is a measure of how far ``off-diagonal''
the decoherence functional is; for large $|\xi|$ it will clearly
be strongly suppressed.  This is ``medium-strength'' decoherence as defined
by Gell-Mann and Hartle \cite{GMHart2}.

Since large $\xi$ is suppressed, we can neglect the higher order terms in
$\xi$ with good accuracy.
This makes the $\xi$ path integral purely quadratic, and therefore
solvable.  Doing this integral yields
\begin{eqnarray}
T(X_1, \xi_1, t_1; && X_0, \xi_0, t_0) = \nonumber\\
  \sqrt{\pi\hbar^2\over{M\gamma kT}} \int \delta X\ \exp\biggl\{
  - {1\over{M\gamma kT}} \int_{t_0}^{t_1}
  \biggl( M {\ddot X}(t) + && dV/dX(X(t))
  + 2M\gamma{\dot X}(t) - g(t) \biggr)^2 dt \nonumber\\
+ (i/\hbar) M (\xi_1 {\dot X}_1 - && \xi_0 {\dot X}_0) \biggr\},
\end{eqnarray}
which is clearly peaked about the solution to the classical equation of
motion
\begin{equation}
M {\ddot X}(t) + dV/dX(X(t)) + 2M\gamma{\dot X}(t) = g(t),
\end{equation}
more and more strongly in the limit of large $M$.

Let $X_{\rm cl}(t)$ be the solution to the above classical equation
with the boundary conditions
$X_{\rm cl}(t_0) = X_0$ and $X_{\rm cl}(t_1) = X_1$.
We can then define a new variable $\eta(t)$
\begin{equation}
\eta(t) = X(t) - X_{\rm cl}(t).
\end{equation}
Clearly $\eta(t)$ has boundary conditions $\eta(t_1) = \eta(t_0) = 0$.
As $M$ becomes large, we can treat $\eta(t)$ as a small deviation, and
approximate the path integral as
\begin{eqnarray}
T(X_1, \xi_1, t_1; && X_0, \xi_0, t_0) = \nonumber\\
\sqrt{\pi\hbar^2\over{M\gamma kT}} \int \delta \eta\ \exp\biggl\{
  - {1\over{M\gamma kT}} \int_{t_0}^{t_1}
  \biggl( M {\ddot\eta}(t) && + d^2V/dX^2(X_{\rm cl}(t))\eta(t)
  + 2M\gamma{\dot\eta}(t) \biggr)^2 dt \nonumber\\
+ (i/\hbar) M (\xi_1 {\dot \eta}_1 - \xi_0 {\dot \eta}_0) &&
  + (i/\hbar) M (\xi_1 {\dot X_{\rm cl}}(t_1)
  - \xi_0 {\dot X_{\rm cl}}(t_0) )\biggr\}.
\label{eta1}
\end{eqnarray}

This path integral is quadratic in $\eta$ and therefore solvable, at least in
principle.  This principle runs into a few problems in practice.  It assumes
that you know $X_{\rm cl}(t)$ as a function of the boundary conditions.
This is true only in very simple cases.  In chaotic cases, it may be difficult
to determine this function even numerically.  Also, this integral contains
(in essence) 4th derivatives of $\eta$, which complicate the calculation
in some ways.  Still, by making a few assumptions about the behavior of
$X_{\rm cl}(t)$, we can still extract some useful information from this
expression.

Since for the purposes of determining decoherence we are really only interested
in the $\xi$ dependence of ${\bf T}$, it is straightforward, albeit tedious,
to show that
\begin{eqnarray}
T(X_1, \xi_1, t_1; && X_0, \xi_0, t_0) =  \nonumber\\
K \exp\biggl\{
  - {{M\gamma kT}\over{\hbar^2}}
   (\lambda_1 \xi_0^2 + \lambda_2 \xi_0 \xi_1
  + \lambda_2 \xi_1^2)
&& + (i/\hbar) M (\xi_1 {\dot X_{\rm cl}}(t_1)
  - \xi_0 {\dot X_{\rm cl}}(t_0) ) \biggr\}.
\label{eta2}
\end{eqnarray}
Note that $\lambda_i = \lambda_i(X_1, X_0, t_1, t_0)$ and
$K = K(X_1, X_0, t_1, t_0)$.  These functions are not especially easy to
calculate, but can be computed numerically if necessary.  Simple calculations
along those lines seem to show that $\lambda_i/(t_1-t_0)$
is relatively constant
for $(t_1 - t_0)$ short compared to the dynamical time of the system and
long compared to the decoherence time, at least for high-probability paths.
For longer times, comparable to
the dynamic timescale of the system in question, the $\lambda_i$ vary
enormously in magnitude; numerical results showed a variability of more than
four orders of magnitude, though most results for $\lambda_1/(t_1 - t_0)$
and $\lambda_3/(t_1 - t_0)$
clustered around certain values, and never became negligibly small.
$\lambda_1 \xi_0^2 + \lambda_2 \xi_0 \xi1 + \lambda_3 \xi_1^2$ is, in any case,
always a strictly non-negative quantity.  For details of these calculations,
see the Appendix.

If we make the (admittedly highly questionable) assumption that the $\lambda_i$
are roughly constant for constant $(t_1-t_0)$,
then we can estimate the level of decoherence achievable
with phase-space projections.  For simplicity, we will only look at the
projections at a single time:
\begin{equation}
{\rm Tr} \biggl\{ \cdots {\bf T}( P^i_{\alpha_i}(t_i)
  {\bf T} ( \cdots ) P^i_{\alpha_i^\prime}(t_i)) \cdots \biggr\}
\end{equation}

\subsection{Consecutive $X$ and $P$ Projections}

Using the approximate $X$ and $P$ projections described in
(\ref{approximate_projections}) above, we can examine decoherence by
looking at the off-diagonal elements, where the projections are centered
on $({\bar x},{\bar p})$ and $({\bar x^\prime},{\bar p^\prime})$
respectively.  A single pair of projections at time $t_i$ will multiply
the decoherence functional by a factor
\begin{eqnarray}
\int dX_1 dX_2 d\xi_1 d\xi_2 dp dp^\prime &&
  T(X_3, \xi_3, t_{i+1}; X_2, \xi_2, t_i) \nonumber\\
  \times \exp\biggl\{ - {{(X_1 + \xi_1/2 - {\bar x})^2}\over{\Delta x^2}} &&
   - {{(X_1 - \xi_1/2 - {\bar x^\prime})^2}\over{\Delta x^2}}
   - {{(p - {\bar p})^2}\over{\Delta p^2}}
   - {{(p^\prime - {\bar p^\prime})^2}\over{\Delta p^2}} \nonumber\\
+ i (p - p^\prime)(X_2 - X_1) + {i\over2} (p + p^\prime) && (\xi_2 - \xi_1)
   \biggr\} \nonumber\\
\times T(X_1, \xi_1, t_i; && X_0, \xi_0, t_{i-1})
\end{eqnarray}

\begin{eqnarray}
\sim && \int dX_1 dX_2
  K^2 \exp\biggl\{ -
  \biggl( { { (X_1 - {\bar x})^2 + (X_1 - {\bar x^\prime})^2
  + (\Delta p^2\Delta x^2/2\hbar^2) (X_2 - X_1)^2 }\over{\Delta x^2}}
  \biggr) \nonumber\\
&& + \biggl( { { (8\hbar^2 M\gamma kT\lambda + \hbar^2\Delta p^2)
  ({\bar x} - {\bar x^\prime})^2 }\over
  { 2\Delta x^2(8\hbar^2 M\gamma kT\lambda + \hbar^2\Delta p^2
  +  16 (M\gamma kT\lambda)^2\Delta x^2
  +  4M\gamma kT\lambda\Delta p^2 \Delta x^2) } } \biggr) \nonumber\\
&& - \biggl( { { (\hbar^2 + 2M\gamma kT\lambda\Delta x^2) \bigl[
  (p_2 - {\bar p})^2 + (p_2 - {\bar p^\prime})^2 \bigr] }\over
  { 8\hbar^2 M\gamma kT\lambda + \hbar^2\Delta p^2
  +  16 (M\gamma kT\lambda)^2\Delta x^2
  +  4M\gamma kT\lambda\Delta p^2 \Delta x^2 } } \biggr) \nonumber\\
&& - \biggl( { { (2M\gamma kT\lambda\Delta x^2) \bigl[
  (p_1 - {\bar p})^2 + (p_1 - {\bar p^\prime})^2 \bigr] }\over
  { 8\hbar^2 M\gamma kT\lambda + \hbar^2\Delta p^2
  +  16 (M\gamma kT\lambda)^2\Delta x^2
  +  4M\gamma kT\lambda\Delta p^2 \Delta x^2 } } \biggr) \nonumber\\
&& + \biggl( { { (\hbar^2 + 4M\gamma kT\lambda\Delta x^2)
  ({\bar p} - {\bar p^\prime})^2 - \Delta p^2 (p_2 - p_1)^2 }\over
  { 2(8\hbar^2 M\gamma kT\lambda + \hbar^2\Delta p^2
  +  16 (M\gamma kT\lambda)^2\Delta x^2
  +  4M\gamma kT\lambda\Delta p^2 \Delta x^2) } } \biggr) \nonumber\\
&& - i \biggl( { { (4\hbar M\gamma kT\lambda) ({\bar x} - {\bar x^\prime})
  ({\bar p} + {\bar p^\prime} - 2 p_1) }\over
  { 8\hbar^2 M\gamma kT\lambda + \hbar^2\Delta p^2
  +  16 (M\gamma kT\lambda)^2\Delta x^2
  +  4M\gamma kT\lambda\Delta p^2 \Delta x^2 } } \biggr) \nonumber\\
&& + {{ i ({\bar p} - {\bar p^\prime})(X_2 - X_1)}\over\hbar} \biggr\},
\end{eqnarray}
where $p_1$ and $p_2$ are $M {\dot X_{\rm cl}}$ for boundary conditions
$\{X(t_{i-1}) = X_0, X(t_i) = X_1\}$
and $\{X(t_i) = X_2, X(t_{i+1}) = X_3\}$
respectively.

A formidable expression indeed!  One can, with difficulty, see that in
general this factor will be suppressed for off-diagonal terms.  If we
simplify matters by taking the semiclassical, high-temperature limit,
an examination of the real terms of the exponent show that for
$|{\bar x} - {\bar x^\prime}|^2 \sim \delta x^2$ this expression is suppressed
by a minimum factor of
\[
\exp\biggl\{ - {{\delta x^2}\over{\Delta x^2}} \biggl({1\over2} -
  {{\hbar^2}\over{4M\gamma kT\lambda\Delta x^2}} \biggr) \biggr\}.
\]

A similar examination of the real $p$ terms gives no similar comfort, for
we find that there the minimum level of suppression is none at all!  This
doesn't mean that histories with differing $p$'s do not decohere;
the last imaginary term in the exponent oscillates extremely rapidly, and
will tend to suppress all off-diagonal terms as $X_1$ and $X_2$ are
integrated over.  This will work, in general if $|{\bar p} - {\bar p^\prime}|^2
\sim \delta p^2$ is large compared to $\Delta p^2$ and $\hbar^2/\Delta x^2$.
To suppress the off-diagonal terms of the
decoherence functional by a factor $\epsilon$, where $\epsilon \ll 1$,
we must have
\begin{eqnarray}
\delta x^2 \ge && - \Delta x^2 \ln \epsilon, \nonumber\\
\delta p^2 \ge && - \Delta p^2 \ln \epsilon, \nonumber\\
\Delta x^2 \Delta p^2 \ge && \hbar^2.
\end{eqnarray}

\subsection{Coherent State Projections}

The results from coherent state projections are similar, but somewhat
cleaner and easier to see.  In this case the factor from the projections
at one time $t_i$ goes as
\begin{eqnarray}
\int dX_1 dX_2 K^2 \exp\biggl\{ &&
  - \biggl( { { (X_1 - {\bar x})^2 + (X_1 - {\bar x^\prime})^2
  + (X_2 - {\bar x})^2 + (X_2 - {\bar x^\prime})^2 }
  \over{2\sigma}} \biggr) \nonumber\\
&& + \biggl( { { \hbar^2 ( {\bar x} - {\bar x^\prime})^2 }\over
  { 2\hbar^2\sigma + 8\hbar M\gamma kT\lambda\sigma^2 }} \biggr) \nonumber\\
&& - \biggl( { { (p_1 - {\bar p})^2 + (p_1 - {\bar p^\prime})^2
  + (p_2 - {\bar p})^2 + (p_2 - {\bar p^\prime})^2
  - ({\bar p} - {\bar p^\prime})^2 }\over
  { 2\hbar^2/\sigma + 8 \hbar M\gamma kT\lambda }} \biggr) \nonumber\\
&& - { { i \hbar ({\bar x} - {\bar x^\prime}) (p_2 - p_1) }\over
  { \hbar^2 + 4 M\gamma kT\lambda\sigma }}
  + { { i ({\bar p} - {\bar p^\prime}) (X_2 - X_1) }\over{\hbar}}
  \biggr\}.
\end{eqnarray}

Here again, we see that in the semiclassical limit this reduces to a
minimal level of suppression
\[
\exp\biggl\{ - {{\delta x^2}\over{\sigma}} \biggl({1\over2} -
  {{\hbar^2}\over{8 M\gamma kT\lambda\sigma}} \biggr) \biggr\}
\]
for the $x$ terms, and that $\delta p^2$ must be large compared to
$\Delta p^2 \ge \hbar^2/\sigma$.

\section{Conclusions}

While no set of approximate phase space projections treated in this paper
is completely satisfactory, they do serve to illustrate certain
traits that phase space histories should possess.  Highly discontinuous
trajectories are suppressed, and as one goes to the semiclassical limit
the probabilities of histories become peaked about the classical solutions.
While precise statements about decoherence are hard to make, given the
difficulty of solving the problem for highly general systems, rough arguments
indicate that the size of phase space cells needed for decoherence is much
larger than that naively indicated by the uncertainty principle
($\Delta x \Delta p \sim \hbar$).

\acknowledgements

The guidance of Murray Gell-Mann is gratefully acknowledged.  I had helpful
conversations and exchanges with many people; among the most notable were
Seth Lloyd, Jim Hartle, Jonathon Halliwell, Carlton Caves, and Jeff Kimble.
Much of the algebra (though not all, alas!) was made easier by use
of the Mathematica symbolic mathematics program.  This work was supported
in part by the U.S. Department of Energy under Grant No. DE-FG03-92-ER40701.

\appendix
\section*{Path Integral for the Transition Matrix}

The path integral in (\ref{eta1}) is somewhat unusual in that it has only
two boundary conditions ($\eta(t_0) = \eta(t_1) = 0$) for an integrand with
four derivatives!  Thus, the usual prescription for solving quadratic path
integrals is not immediately applicable.

This procedure can still be used, however, by the simple expedient of
imposing two more boundary conditions, ${\dot \eta}(t_0) = v_0$ and
${\dot \eta}(t_1) = v_1$, and solving the path integral, then integrating
the result over all values of $v_0$ and $v_1$.

The path integral to be solved is then
\begin{eqnarray}
\int \delta\eta\ \exp\biggl\{ S[\eta(t)]
  + && i (M/\hbar)(\xi_1 v_1 - \xi_0 v_0) \biggr\} = \nonumber\\
&& F(t_1,t_0) \exp\bigl\{ S_{\rm cl}
  + i (M/\hbar)(\xi_1 v_1 - \xi_0 v_0) \bigr\},
\end{eqnarray}
where
\begin{equation}
S[\eta(t)] = - {1\over{M\gamma kT}} \int_{t_0}^{t_1} \varepsilon^2(t) dt,
\end{equation}
\begin{equation}
\varepsilon(t) = M ({\ddot \eta} + 2\gamma{\dot \eta} + f(t)\eta ),
\label{e_of_t}
\end{equation}
\begin{equation}
f(t) = {1\over M}{{d^2V}\over{dX^2}}(X_{\rm cl}(t)),
\end{equation}
and $F(t_1,t_0)$ is an undetermined multiplier independent of
the boundary conditions of $\eta$.  (Of course, since $f(t)$ is defined
in terms of $X_{\rm cl}(t)$, this whole solution {\it is} dependent on
$X_0$ and $X_1$.  This dependence is complicated, as we will see.)

The classical action $S_{\rm cl}$ is the action of the path $\eta(t)$ that
obeys the classical equation of motion.  For
\[
S = \int_{t_0}^{t_1} L(\eta,{\dot \eta},{\ddot \eta}) dt
\]
the Euler-Lagrange equation is
\begin{equation}
{{d^2}\over{dt^2}}\left( {{\partial L}\over{\partial {\ddot \eta}}} \right)
  - {d\over{dt}} \left( {{\partial L}\over{\partial {\dot \eta}}} \right)
  + \left( {{\partial L}\over{\partial \eta}} \right) = 0.
\label{euler_lagrange}
\end{equation}
Plugging in our definitions for $S[\eta(t)]$ and $\varepsilon(t)$,
we get the equations
\begin{mathletters}
\begin{eqnarray}
{\ddot \eta} + 2\gamma{\dot\eta} + f(t) \eta && = \varepsilon(t)/M, \\
{\ddot \varepsilon} - 2\gamma{\dot \varepsilon} + f(t) \varepsilon && = 0.
\end{eqnarray}
\end{mathletters}
The first of these follows from the definition (\ref{e_of_t}), the
second from (\ref{euler_lagrange}).

The solution to the first equation is
\begin{equation}
\varepsilon(t) = M {\rm e}^{\gamma t} (A a_1(t) + B a_2(t)),
\end{equation}
where $a_1$ and $a_2$ are two independent solutions of the equation
\begin{equation}
{\ddot a} = (\gamma^2 - f(t)) a.
\label{a_eqn}
\end{equation}
This equation is not easily solved analytically in most cases.  For slowly
varying $f(t)$ one can approximate the solution
\begin{equation}
a(t) \approx \exp\biggl\{ \pm \int_{t_0}^t \sqrt{\gamma^2 - f(s)} ds \biggr\};
\end{equation}
in any case, (\ref{a_eqn}) is readily solvable numerically.  While any
independent boundary conditions will work for $a_1$ and $a_2$, a convenient
choice is
\begin{eqnarray}
a_1(0) = 1, && {\dot {a_1}}(0) = 0; \nonumber\\
a_2(0) = 0, && {\dot {a_2}}(0) = 1.
\end{eqnarray}

This is then plugged into the equation for $\eta$ to give the solution
\begin{eqnarray}
\eta(t) = && {\rm e}^{- \gamma t}( C a_1(t) + D a_2(t) ) \nonumber\\
&&  + ({\rm e}^{- \gamma t}/M) \int_{t_0}^t \biggl(
  {{a_1(t) a_2(s) - a_2(t) a_1(s)}
  \over{{\dot {a_1}}(s) a_2(s) - {\dot{a_2}}(s) a_1(s)}} \biggr)
  \varepsilon(s) {\rm e}^{2 \gamma s} ds.
\end{eqnarray}
Imposing the boundary conditions on $\eta$ then gives us a set of equations
involving $A$, $B$, $C$, and $D$:
\begin{mathletters}
\begin{eqnarray}
C = && 0, \\
D = && v_0, \\
v_0 a_2(t_1) + A (a_1(t_1) I_2 - a_2(t_1) I_1) &&
  + B (a_1(t_1) I_3 - a_2(t_1) I_2) = 0, \\
v_0 {\dot{a_2}}(t_1) + A ({\dot{a_1}}(t_1) I_2 - {\dot{a_2}}(t_1) I_1) &&
  + B ({\dot{a_1}}(t_1) I_3 - {\dot{a_2}}(t_1) I_2)
  = v_1 {\rm e}^{\gamma t_1},
\end{eqnarray}
\end{mathletters}
where
\begin{eqnarray}
I_1 && = \int_{t_0}^{t_1} {{a_1(t)^2 {\rm e}^{2 \gamma t} dt}\over
  {{\dot{a_1}}(t) a_2(t) - {\dot{a_2}}(t) a_1(t)}},\\
I_2 && = \int_{t_0}^{t_1} {{a_1(t) a_2(t) {\rm e}^{2 \gamma t} dt}\over
  {{\dot{a_1}}(t) a_2(t) - {\dot{a_2}}(t) a_1(t)}},\\
I_3 && = \int_{t_0}^{t_1} {{a_2(t)^2 {\rm e}^{2 \gamma t} dt}\over
  {{\dot{a_1}}(t) a_2(t) - {\dot{a_2}}(t) a_1(t)}}.
\end{eqnarray}
Again, these integrals are usually only solvable numerically.

Solving for $A$ and $B$ then yields
\begin{eqnarray}
A && = {{ ( {\dot a_1}(t_1) a_2(t_1) - a_1(t_1) {\dot a_2}(t_1) ) I_3 v_0
  + {\rm e}^{\gamma t_1} (a_1(t_1) I_3 - a_2(t_1) I_2) v_1 }\over
  { (a_1(t_1) {\dot a_2}(t_1) - {\dot a_1}(t_1) a_2(t_1))
  (I_2^2 - I_1 I_3) }} \nonumber\\
&& = A_0 v_0 + A_1 v_1,
\end{eqnarray}
\begin{eqnarray}
B && = {{ - ( {\dot a_1}(t_1) a_2(t_1) - a_1(t_1) {\dot a_2}(t_1) ) I_2 v_0
  - {\rm e}^{\gamma t_1} (a_1(t_1) I_2 - a_2(t_1) I_1) v_1 }\over
  { (a_1(t_1) {\dot a_2}(t_1) - {\dot a_1}(t_1) a_2(t_1))
  (I_2^2 - I_1 I_3) }} \nonumber\\
&& = B_0 v_0 + B_1 v_1.
\end{eqnarray}
Plugging these results back into the definitions of $\varepsilon(t)$ and
$S[\eta(t)]$ gives a value for the path integral
\begin{eqnarray}
\exp\biggl\{ S_{\rm cl}
&& + (i M/\hbar)(\xi_1 v_1 - \xi_0 v_0) \biggr\} = \nonumber\\
&& \exp\biggl\{ - {M\over{\gamma kT}}
  \bigl( A^2 I_4 + 2 A B I_5 + B^2 I_6 \bigr)
  + i (M/\hbar)(\xi_1 v_1 - \xi_0 v_0) \biggr\},
\end{eqnarray}
where
\begin{eqnarray}
I_4 && = \int_{t_0}^{t_1} a_1(t)^2 {\rm e}^{2 \gamma t} dt,\\
I_5 && = \int_{t_0}^{t_1} a_1(t) a_2(t) {\rm e}^{2 \gamma t} dt,\\
I_6 && = \int_{t_0}^{t_1} a_2(t)^2 {\rm e}^{2 \gamma t} dt.
\end{eqnarray}

Clearly, the exponent is quadratic in $v_0$ and $v_1$.  Integrating over
these two boundary conditions, we perform two gaussian integrals, and arrive
at our final result
\begin{eqnarray}
\int dv_0 dv_1 \int \delta\eta\ \exp\biggl\{ S[\eta(t)]
&& + i (M/\hbar)(\xi_1 v_1 - \xi_0 v_0) \biggr\} = \nonumber\\
&& K \exp\biggl\{ - {{M\gamma kT}\over{\hbar^2}}
  \bigl( \lambda_1 \xi_0^2 + \lambda_2 \xi_0 \xi_1
  + \lambda_3 \xi_1^2 \bigr) \biggr\}.
\end{eqnarray}
where
\begin{mathletters}
\begin{equation}
\lambda_1 = { { A_1^2 I_4 + 2 A_1 B_1 I_5 + B_1^2 I_6 }\over
  { 4 (A_0 B_1 - A_1 B_0)^2 (I_5^2 - I_4 I_6) } },
\end{equation}
\begin{equation}
\lambda_2 = { { A_0 A_1 I_4 + (A_1 B_0 + A_0 B_1) I_5 + B_0 B_1 I_6 }\over
  { 2 (A_0 B_1 - A_1 B_0)^2 (I_5^2 - I_4 I_6) } },
\end{equation}
\begin{equation}
\lambda_3 = { { A_0^2 I_4 + 2 A_0 B_0 I_5 + B_0^2 I_6 }\over
  { 4 (A_0 B_1 - A_1 B_0)^2 (I_5^2 - I_4 I_6) } },
\end{equation}
\begin{equation}
K = { { \pi \gamma kT }\over
  { M \sqrt{ (A_1 B_0 - A_0 B_1)^2 (I_4 I_6 - I_5^2) } } } F(t_1, t_0).
\end{equation}
\end{mathletters}

Clearly, these quantities are dependent on $X_0$ and $X_1$:
$\lambda_i = \lambda_i(X_0, X_1, t_0, t_1)$ and
$K = K(X_0, X_1, t_0, t_1)$.  In principle,
$K$ can be calculated, but in practice it is not necessary to do so in order
to make arguments about decoherence.  It would be necessary to do so
in order to actually
compute the probability of a history.

The derivations in this section, tedious as they are, can nevertheless be
readily automated.  Once one determines the classical trajectory $X_{\rm cl}$
which corresponds to the boundary conditions $X_0$ and $X_1$, determining
the $\lambda_i$ numerically is straightforward.  I have used this technique
to examine the values of the $\lambda_i$ for the forced, damped duffing
oscillator model.  For long times $(t_1 - t_0)$ the $\lambda_i$ varied
enormously in magnitude as a function of $X_0$ and $X_1$.  For times short
compared to the dynamical time of the system, however, the $\lambda_i$ were
nearly linear:  $\lambda/(t_1-t_0) \approx 0.085$.  For some boundary
conditions
they might become considerably larger, but in what was, admittedly, not an
exhaustive sampling, none got much smaller.

\end{document}